\documentstyle[12pt,epsfig]{article}

\newcommand{\bce}{\begin{center}}
\newcommand{\ece}{\end{center}}
\newcommand{\beq}{\begin{equation}}
\newcommand{\eeq}{\end{equation}}
\newcommand{\be}{\begin{equation}}
\newcommand{\ee}{\end{equation}}
\newcommand{\bea}{\vspace{0.25cm}\begin{eqnarray}}
\newcommand{\eea}{\end{eqnarray}}

\newcommand{\ba}{\begin{array}}
\newcommand{\ea}{\end{array}}
\newcommand{\doublespace}{
    \renewcommand{\baselinestretch}{1.6}\large\normalsize}
\def\lsim{\mathrel{\rlap{\lower4pt\hbox{\hskip1pt$\sim$}}
    \raise1pt\hbox{$<$}}}         %less than or approx. symbol
\def\gsim{\mathrel{\rlap{\lower4pt\hbox{\hskip1pt$\sim$}}
    \raise1pt\hbox{$>$}}}         %greater than or approx. symbol

\def\Pom{{\bf I\!P}}

 \advance\oddsidemargin   by -1in
 \advance\evensidemargin  by -1in
 \advance\topmargin   by -0.5in
\begin{document}

\phantom{.}{\large \bf \hspace{10.2cm}  ISN 97.99 \\
\phantom{.} \bf \hspace{10.2cm} DFTT 46/97
\vspace{.4cm}\\ }

\begin{center}
{\Large \bf Twist-4 effects and $Q^{2}$ dependence of diffractive
DIS
\vspace{1.0cm}\\}
{\large \bf M. Bertini$^{a}$, M. Genovese\footnote{ \small
Supported by the EU Program ERBFMBICT 950427} $^{b}$, N.N.~Nikolaev$^{c,d}$,
A.V.Pronyaev$^{e}$,
B.G.~Zakharov$^{d}$ \vspace{1.0cm}\\}
{\it

$^{a}$ INFN, Sezione di Torino, Via P.Giuria 1, I-10125 Torino, Italy
\medskip \\
$^{b}$  Institut des Sciences Nucl\'eaires \\
Universit\'e Joseph Fourier--IN2P3-CNRS, \\
53, avenue des Martyrs, F-38026 Grenoble Cedex,
France
\medskip \\
$^{c}$IKP(Theorie), FZ J{\"u}lich, 5170 J{\"u}lich, Germany
\medskip\\
$^{d}$ L. D. Landau Institute for Theoretical Physics, GSP-1,
117940, \\
ul. Kosygina 2, Moscow V-334, Russia.
\medskip\\
$^{d}$Department of Physics,
Virginia Polytechnic Institute and State University,\\
Blacksburg, VA 24061, USA
\vspace{1cm}\\}

{\Large \bf Abstract} \\
\end{center}
%marco changes following
In this letter we report the direct perturbative QCD evaluation of twist-4
effects in diffractive DIS. They are large and have a strong
impact on the $Q^{2}$ dependence of diffractive structure functions
at large $\beta$.  Based on the
AGK rules, we comment on the possible contribution from diffractive
higher twists to $\propto {1 \over Q^{2}}$ corrections to proton
structure function at small $x$. These corrections to the longitudinal
structure function $F_{L}$ may be particularly large.
\vspace{2cm}

\begin{center}
E-mail: kph154@zam001.zam.kfa-juelich.de
\end{center}

\newpage
%================================================================================

\doublespace

The direct calculation of higher twist corrections to the proton structure
function (SF) remains one of challenges of perturbative QCD (pQCD) description
of inclusive deep inelastic scattering (DIS) (\cite{Shuryak,Jaffe},
for phenomenological estimates based on departures from the DGLAP
evolution see \cite{Phenom}). In a striking contrast to inclusive
proton SF, twist-4 (T4) corrections to diffractive SF,
$F_{i}^{D(3)}(x_{\Pom},\beta,Q^{2})$, $(i=L,T)$, are genuinely short
distance dominated \cite{NZ91,NZ92} and thus  calculable
from the first principles of pQCD. Furthermore, they are strongly
enhanced by the gluon SF of the proton squared
\cite{GNZlong}, $F_{i}^{D(3)}(T4;x_{\Pom},\beta,Q^{2}) \propto
{1\over Q^{2}}G^{2}(x_{\Pom},Q^{2})$. (Hereafter $x, Q^{2}, \beta$ and
$x_{\Pom}={x\over \beta}$ are standard diffractive DIS variables and
$G(x,Q^2)= x g(x,Q^2)$).
 For instance, for $\beta\gsim 0.9$,
the twist-4 longitudinal diffractive SF was found to be
comparable to, and even exceeding, the leading twist-2 (T2) transverse
SF over a broad range of $Q^{2}$, making questionable
\cite{GNZlong,GNZcharm,NZDIS96,NZDIS97} applications of the standard DGLAP
evolution to diffractive SF.

In this letter we extend our early analysis of twist-4 longitudinal
diffractive SF \cite{GNZlong} to similarly enhanced twist-4 contribution
to the large-$\beta$ transverse diffractive SF
\footnote{The preliminary results from
this study have been reported at DIS'97 \cite{NZDIS97}. Related
results were presented at DIS'97 by J.Bartels and M.W\"usthoff
\cite{BartelsDIS97}.}. It enters  $F_{T}^{D(3)}(x_{\Pom},\beta,Q^{2})$
with negative sign and, as the higher twist dies out at large $Q^{2}$, the
transverse SF $F_{T}^{D(3)}(x_{\Pom},\beta,Q^{2})$ will rise with
increasing $Q^{2}$. We present for the first time an analysis of the
$\beta$ and $Q^{2}$ dependence of compensations of the longitudinal
and transverse twist-4 SF's in $F_{2}^{D(3)}=F_{T}^{D(3)}+F_{L}^{D(3)}$,
which completes the evaluation of driving twist-4 terms in $F_{2}^{D(3)}$.
We compare our results with the H1 experimental data on $F_{2}^{D(3)}$
\cite{H1F2D3}. Based on a connection between diffractive DIS and unitarity
corrections to the proton SF suggested by the Abramovski-Gribov-Kancheli
(AGK) rules \cite{AGK,BaroneUnit,NZ94}, we evaluate the unitarity-driven
diffractive twist-4 corrections to the small-$x$ proton SF, which is
potentially very large in the proton longitudinal SF $F_{L}$.

The underlying pQCD subprocess for diffractive DIS at large $\beta$
is the excitation of the $q\bar{q}$ Fock states of the photon
$\gamma^{*}p\rightarrow Xp'$, where $X=q\bar{q}$ \cite{NZ92}. The relevant
pQCD diagrams are shown in Fig.~1. In what follows, $z$ and $(1-z)$ are
the fractions of the (light--cone) momentum of the photon carried
by the quark and
antiquark, respectively, $\vec{k}$ is the relative transverse momentum in
the $q\bar{q}$ pair, $M^2=(m_{f}^2 + k^2)/z(1-z)$
is the invariant mass of the diffractive system, $\vec{p}_{\perp}$ is the
(p,p') momentum transfer and $t=-{\vec{p}_{\perp}}\,^2$.
The contribution of excitation of one open flavour of electric charge
$e_{f}$ (in units of  the electron charge) and mass $m_{f}$ to
transverse (T) and longitudinal (L) diffractive SF's for forward
diffraction dissociation is calculable in terms of the helicity
amplitudes $\vec{\Phi}_{1}$ and $\Phi_{2}$ introduced in \cite{NZ92}:
\bea
F_{T}^{D(4)}(t=0,x_{\Pom},\beta,Q^{2})=%\nonumber\\
{4\pi e_{f}^{2}\beta  \over 3\sigma_{tot}(pp)}
\int {dk^{2} (k^{2}+m_{f}^{2})\over (1-\beta)^{2}J}
\alpha_{S}^{2}(\bar{Q}^{2})
\left\{\left[1-2{k^{2}+m_{f}^{2} \over M^{2}}\right] \vec{\Phi}_{1}^{2}  +
m_{f}^{2}\Phi_{2}^{2}
\right\} \, ,
\label{eq:FT}
\eea
\bea
F_{L}^{D(4)}(t=0,x_{\Pom},\beta,Q^{2})={4\pi e_{f}^{2} \beta
\over 3\sigma_{tot}(pp)}
\int
{dk^{2} (k^{2}+m_{f}^{2}) \over (1-\beta)^{2}J}
\alpha_{S}^{2}(\bar{Q}^{2})
4z^{2}(1-z)^{2}Q^{2}\Phi_{2}^{2}\,,
\label{eq:FL}
\eea
where $J=\sqrt{1-4{m_{f}^{2}+k^{2}\over M^{2}} }$ is the Jacobian peak factor
\footnote{We follow the convention in which diffractive structure
functions $F_{i}^{D(4)}$ are dimensionless  \cite{NZDIS97}, hence
$\sigma_{tot}(pp)=40mb$ in the denominator
of the {\sl r.h.s} of Eqs. (\ref{eq:FT}), (\ref{eq:FL}). Our definition
(\ref{eq:FD3}) of $F_{i}^{D(3)}$ differs from \cite{H1F2D3,ZEUSF2D4} by
the factor $x_{\Pom}$, so that our $F_{i}^{D(3)}$ does not blow up
$\sim {1\over x_{\Pom}}$ when $x_{\Pom}\rightarrow 0$.}. To the Leading
Log${1\over x}$ approximation \cite{NZsplit},
\bea
\vec{\Phi}_{1} &=&\vec{k} \int \frac{d\kappa^{2}}{\kappa^{4}}
f(x_{\Pom},\kappa^{2})
\left[{1\over k^{2}+\varepsilon^{2}}-{1\over \sqrt{a^{2}-b^{2}}}+
{2\kappa^{2} \over a^{2}-b^{2}+a\sqrt{a^{2}-b^{2}}}\right]\nonumber\\
&\ & \approx 2\vec{k} (1-\beta)^{2}G(x_{\Pom},\bar{Q}^{2})
\left[
{\beta \over (k^{2}+m_{f}^{2})^{2}}+ {(1-\beta)m_{f}^{2}
\over (k^{2}+m_{f}^{2})^{3}}\right]
\, ,
\label{eq:phi1}
\nonumber\\
\nonumber\\
\Phi_{2}&=&\int \frac{d\kappa^{2}}{\kappa^{4}}
f(x_{\Pom},\kappa^{2})
\left[{1\over \sqrt{a^{2}-b^{2}}} -
{1\over k^{2}+\varepsilon^{2}}\right] \nonumber\\
&\ & \approx -(1-\beta)^{2}G(x_{\Pom},\bar{Q}^{2})
\left[
{2\beta -1\over (k^{2}+m_{f}^{2})^{2}}+ {2(1-\beta)m_{f}^{2}
\over (k^{2}+m_{f}^{2})^{3}}\right]
\, .
\label{eq:phi2}
\eea

\noindent
Here $f(x,\kappa^{2})=\partial G(x,k^{2})/\partial \log(\kappa^{2})$
is the unintegrated gluon SF of the proton,
$\varepsilon^{2}=z (1-z) Q^2 + m_{f}^2,~a=\varepsilon^{2}+
k^{2}+\kappa^{2}$, $b=2k\kappa$,
\beq
\bar{Q}^{2} = \varepsilon^{2}+k^{2}={k^{2} + m_{f}^{2} \over 1-\beta}
\label{eq:Q2scale}
\eeq
is the pQCD scale of the Leading Log$Q^{2}$ approximation
(\cite{GNZcharm,NZsplit}), and we also have shown the Leading
Log$Q^{2}$ approximations for $\vec{\Phi}_{1}, \Phi_{2}$. $\bar{Q}^{2}$
has already been used as such in (\ref{eq:FT}), (\ref{eq:FL}) as the
argument of the
strong coupling $\alpha_{S}$. The above outlined formalism has become
a standard description of DIS at large $\beta$
\cite{GNZlong,GNZcharm,GNZ95,Bartels,Levin}. Extra
contributions, $\propto f(x_{\Pom},\bar{Q}^{2})$, which take over in
$\vec{\Phi}_{1},\Phi_{2}$ at $\beta \ll 1$, were found in \cite{NZsplit},
but they can be neglected at $\beta \sim 1$.

The advantage of forward diffraction is that the SF
$F_{i}^{D(4)}(t=0,x_{\Pom},\beta,Q^{2})$ is calculable directly in terms
of the conventional gluon SF. The first measurements of the $t$-dependent
$F_{i}^{D(4)}(t,x_{\Pom},\beta,Q^{2})$ have been reported recently
by the ZEUS collaboration \cite{ZEUSF2D4}, but for large
$\beta$ of our interest in this study only the $t$-integrated diffractive
SF's are available \cite{H1F2D3}.  Assuming
the usual
$d\sigma/dt \propto \exp(-B_{d}\,p_{\perp}^{2})$, one finds
\be
F_{i}^{D(3)}(x_{\Pom},\beta,Q^{2})
=\int dt {\sigma_{tot}(pp) \over 16\pi}
F_{i}^{D(4)}(t,x_{\Pom},\beta,Q^{2}) \approx
{\sigma_{tot}(pp) \over 16\pi B_{d}(\beta)}
F_{i}^{D(4)}(t=0,x_{\Pom},\beta,Q^{2})\,.
\label{eq:FD3}
\ee

Using Eqs. (\ref{eq:FL}),(\ref{eq:phi2}), one readily finds that
the longitudinal SF $F_{L}^{D(4)}$ is dominated
by the contribution from large $k^{2}\sim {1\over 4}M^{2}$, which is
further enhanced because of scaling violations in
the gluon structure function of the proton.
Consequently, it is perturbatively calculable \cite{NZ91,NZ92,GNZlong}
and at large $Q^{2}$ one finds the higher twist
\footnote{We emphasize that in this paper we focus
on $\beta\sim 1$. As it has been shown in \cite{GNZ95}, at $\beta \ll 1$ ,
dominated by the $q\bar{q}g$ excitation, the longitudinal diffractive
SF is twist-2 and
$R=\sigma_{L}/\sigma_{T} \approx 0.2$, in close similarity to
inclusive DIS \cite{NZHERA}.}.
\bea
F_{L}^{D(3)}(T4;t=0,x_{\Pom},\beta,Q^{2})=
{4\beta^{3}(1-2\beta)^{2}\over Q^{2}B_{3\Pom}}\cdot { e_{f}^{2}
\over 12}{\cal F_{G}}
(x_{\Pom},\beta,Q^{2})\,,
\label{eq:FLtwist4}
\eea
where
\be
{\cal F_{G}}
(x_{\Pom},\beta,Q^{2})=\int\limits_{{m_{f}^{2} \over 1-\beta}}
^{{Q^{2} \over 4\beta}}
{d\bar{Q}^{2} \over \bar{Q}^{2}}
\left(1-{4\beta \bar{Q}^{2} \over Q^{2}}\right)^{-{1\over 2}}
\left[\alpha_{s}(\bar{Q}^{2})G(x_{\Pom},\bar{Q}^{2})\right]^{2}\,,
\label{eq:Gsqr}
\ee
and for quick estimates to a logarithmic accuracy,
${\cal F_{G}}(x_{\Pom},\beta,Q^{2}) \approx [\alpha_{S}({1\over 4}Q^{2})
G(x_{\Pom},{1\over 4}Q^{2})]^{2}$.

For the separation of the transverse SF $F_{T}^{D(4)}$ into the twist-2
and twist-4 components, we notice that
\beq
{k^{2}+m_{f}^{2} \over M^{2}}=
{\beta \over 1-\beta}\cdot {k^{2}+m_{f}^{2} \over
Q^{2} }\,,
\label{eq:zfactor}
\eeq
which defines the decomposition
\bea
F_{T}^{D(3)}(T2;t=0,x_{\Pom},\beta,Q^{2})=
{ e_{f}^{2} \over 12B_{d}(T2;\beta)}
\int {dk^{2} (k^{2}+m_{f}^{2})\beta \over J(1-\beta)^{2}}
\alpha_{S}^{2}(\bar{Q}^{2})
\left\{ \vec{\Phi}_{1}^{2}  +
m_{f}^{2}\Phi_{2}^{2}
\right\} \, ,
\label{eq:FT2}
\eea
\bea
F_{T}^{D(3)}(T4;t=0,x_{\Pom},\beta,Q^{2})=-
{e_{f}^{2}\beta \over 6(1-\beta)Q^{2}B_{3\Pom}}
\int {dk^{2} (k^{2}+m_{f}^{2})^{2}\beta \over J(1-\beta)^{2}}\cdot
 \alpha_{S}^{2}(\bar{Q}^{2})
\vec{\Phi}_{1}^{2} \, .
\label{eq:FT4}
\eea
Notice, that the large-$k^{2}$ behaviour of the integrand of the
the twist-4 transverse SF $F_{T}^{D(3)}(T4)$ is identical to that
of $F_{L}^{D(3)}$, it is pQCD calculable, and for large $Q^{2}$, well
above the charm threshold, we find
\bea
F_{T}^{D(3)}(T4;t=0,x_{\Pom},\beta,Q^{2})=
-{8 \beta^{4}(1-\beta)\over Q^{2} B_{3\Pom}}\cdot { e_{f}^{2}
\over 12}
{\cal F_{G}}
(x_{\Pom},\beta,Q^{2})\,.
\label{eq:FTtwist4}
\eea
In contrast to that, the
twist-2 transverse SF is dominated by $k^{2}\sim m_{f}^{2}$ :
\bea
F_{T}^{D(4)}(T2;t=0,x_{\Pom},\beta,Q^{2}) \propto
\int{dk^{2} k^{2} \over (k^{2}+m_{f}^{2})^{3}}\cdot
\left[\alpha_{S}(\bar{Q}^{2})G(x_{\Pom},\bar{Q}^{2} =
{m_{f}^{2}+k^{2} \over (1-\beta)})\right]^{2}\, ,
\label{eq:FTk2}
\eea
(we suppress in (\ref{eq:FTk2}) terms containing extra factors
$[m_{f}^{2}/(k^{2}+m_{f}^{2})]^{n}$). Still,
for heavy flavours and/or sufficiently small $(1-\beta)$, the pQCD scale
$\bar{Q}^{2}$ is large, one is in the legitimate pQCD domain and to
a logarithmic accuracy,
\bea
F_{T}^{D(3)}(T2;x_{\Pom},\beta,Q^{2})\approx %\nonumber\\
{\beta (1-\beta)^{2}(3+4\beta+8\beta^2)\over 6 m_{f}^{2}B_{d}(T2;\beta)}
\cdot
{ e_{f}^{2} \over 12}
\cdot\left[\alpha_{S}(\bar{Q}^{2})
G^{2}(x_{\Pom},\bar{Q}^{2}) \approx {m_{f}^{2} \over (1-\beta)})\right]^{2}\, .
\label{eq:FTtwist2}
\eea
Both terms, $\propto \vec{\Phi}_{1}^2$ and $\propto \Phi_{2}^2$
in Eq.~(\ref{eq:FT2}),
give comparable contributions to the twist-2 $F_{T}^{D(4)}$.
Because of the finite phase space of the $k^{2}$ integration, $k^{2}
\leq {1\over 4}M^{2}-m_{f}^{2}$, there emerge higher-twist looking small
corrections $\propto \left({m_{f}^{2}\over M^{2}}\right)^{n}, {n\geq 1},$
to the above defined twist-2 SF. These corrections are logarithmically
weaker than the genuine twist-4 transverse SF (\ref{eq:FTtwist4}) and
for this reason, we do not separate them.

Evaluating the twist-2 transverse SF at not so large $\beta$, one
needs a model for the small-$Q^2$ behaviour of the unintegrated
gluon structure function $f(x,Q^{2})$. Because of cancelation
of soft gluon radiation from different quarks in the
colour singlet nucleon, $f(x,Q^{2})$ vanishes at $Q^{2}\rightarrow 0$.
The Born approximation suggests the soft $Q^{2}$ behaviour
\cite{NZsplit,Baroneglue}
\beq
{\partial G_{B}(Q^{2})\over \partial \log Q^{2}} =
{4C_{B}\alpha_{S}(Q^{2}) \over \pi} {Q^{4} \over
(Q^{2}+\mu_{G}^{2})^{2}}[1-G_{2}(Q,-Q)]\, ,
\label{eq:GBorn}
\eeq
where  $G_{2}(Q,-Q)$ is the two-quark form factor of the nucleus,
$G(0,0)=1$. For harder gluons one
must interpolate between the form (\ref{eq:GBorn}) and any convenient
parameterization for $G(x,Q^{2})$, for instance, the GRV NLO
parameterization \cite{GRV}:
\beq
G(x,Q^{2})=G_{B}(Q^{2})\left({Q_{0}^{2} \over Q_{0}^{2} +Q^{2}}\right)^N+
G_{GRV}(x,Q^{2})\left[1-\left({Q_{0}^{2} \over Q_{0}^{2} +Q^{2}}\right)^N
\right]
\label{Ginterp}
\eeq
For a similar soft $Q^{2}$ parameterizations see Ref. \cite{Bartels}.
The GRV formulas hold only for $Q^{2} \geq  Q_{c}^{2}=0.4$\,GeV$^{2}$,
we take $G_{GRV}(x,Q^{2}<Q_{c}^{2})=G_{GRV}(x,Q_{c}^{2})$. We
choose the parameters $C_{B}=1.5,~ \mu_{G}=0.15 {\rm GeV},~ Q_{0}^{2}=
3 {\rm GeV}^{2},~ N=1$
so as to reproduce the colour dipole cross section \cite{NNZscan}. This
way, our results for diffractive DIS are very close to those found in
the colour dipole model \cite{GNZ95,GNZcharm}. For the quark masses we
take $m_c=1.5$ GeV, $m_s=0.3$ GeV and $m_{u,d}= 150$ MeV as in
\cite{GNZ95}. For light flavours, Eq. (\ref{eq:FTtwist2}) must not be
taken literary, in this case the pQCD scale $\bar{Q}^{2}$ is set not
by $m_{f}^{2}$ but rather by a range of rapid variation of the gluon
structure function, $\bar{Q}^{2} \sim 0.5$ GeV$^{2}$.

%marco changed in the following paragraph
In order to calculate $F^{D (3)}_i$ we need to know the diffraction
slope $B_{d}$.
It has been anticipated some time ago \cite{NZ91,NZ92,NZ94,GNZ95} that the
typical scale for the diffraction slope $B_{d}(\beta)$ is set by
$B_{3\Pom}\approx$ 6 GeV$^{-2}$, as seen in hadronic diffraction in
the so-called triple pomeron region, which in DIS corresponds to
 $\beta\ll 1$.
This prediction has been confirmed by the first data from the ZEUS LPS:
$B_{d} = 7.1\pm 1.1^{+07}_{-1.0}$ GeV$^{-2}$ in diffractive DIS for
5 GeV$^{-2} <Q^{2}< 20$ GeV$^{-2}$ \cite{Grothe} and $B_{d}=7.7\pm
0.9 \pm 1.0$ GeV$^{-2}$ in real photoproduction \cite{Briskin}.
For the  numerical calculation  of the twist-2 transverse SF
we will use the results of
Ref. \cite{NZDIS97,NPZslope}, where it has been calculated that
$B_{d}(T2)\approx B_{3\Pom}$ for $\beta \rightarrow 0$,
it rises by $\approx 50 \%$ reaching
a maximum at $\beta \sim 0.5$, then drops back to $B_{3\Pom}$ at
$\beta\sim 0.9$, has a sharp minimum at $\beta \rightarrow 1$ and
ends up at $B_{d}(\beta=1) \approx B_{3\Pom}$ in the exclusive
limit of vector meson  production. For the short-distance dominated
twist-4 SF's only the proton size contributes to the diffraction slope
$B_{d}$ and, consequently, we will use
$B_{d}(T4;\beta) \approx B_{3\Pom}$ \cite{NZDIS97,NPZslope}.

%marco changes in the following paragraph
For comparing the large-$\beta$ behaviour of the twist-4 and twist-2
contributions to the transverse SF at large $\beta$, it is
sufficient to use the logarithmic approximation, in which 
\bea
{F_{T}^{D(3)}(T4;t=0,x_{\Pom},\beta,Q^{2})
\over
F_{T}^{D(3)}(T2;t=0,x_{\Pom},\beta,Q^{2})}
\approx 
\qquad \qquad ~~~~~~~~~~~~~~~~~~~~~~~~~~~~~~~~~~~~~~~~~~~~~~\nonumber\\
-{48  m_{f}^{2} \over (3+4\beta+8\beta^2)Q^{2}}
\cdot{B_{d}(T2;\beta)\over B_{3\Pom}}
\cdot
{\beta^{3} \over 1-\beta}\cdot
\left[\alpha_{S}({1\over 4}Q^{2})G(x_{\Pom},{1\over 4}Q^{2}) \over
{\alpha_{S}(\bar{Q}^{2})G(x_{\Pom},\bar{Q}^{2}\approx
{m_{f}^{2} \over (1-\beta)})}\right]^{2}\, .
\label{Ratio}
\eea
The rise of this ratio $\propto {1\over 1-\beta}$ when $\beta
\rightarrow 1$ is to a large extent compensated  by the growth of
$G^{2}(x_{\Pom},\bar{Q}^{2})$ because of the rising pQCD scale
$\bar{Q}^{2}\propto {1\over 1-\beta}$ and, as a minor effect for $\beta
> 0.9$, by the decrease of $B_{d}(T2;\beta)$ with
$\beta \rightarrow 1$ as it was found in \cite{NPZslope}.

If the longitudinal polarization
of the virtual photon equals $\epsilon_{L}$, then the twist-4 T and L
SF's enter in the combination $F_{2\epsilon_{L}}=F_{T}+
\epsilon_{L} F_{L}$. To the leading Log$Q^{2}$ approximation,
one readily finds
\bea
F_{2\epsilon_{L}}^{D(3)}(T4;x_{\Pom},\beta,Q^{2})= { e_{f}^{2}
 \over
6 Q^{2}B_{3\Pom}}
\beta^{3}[(1+2\epsilon_{L})(1-2\beta)^{2}-1]
{\cal F_{G}}
(x_{\Pom},\beta,Q^{2})\,,
\label{eq:F2twist4}
\eea
and the overall twist-4 contribution is dominated by the the positive
valued $F_{L}^{D(3)}$ for $\beta$ beyond the crossover points,
$\beta >\beta_{+}$ or $\beta < \beta_{-}$,
and by the  negative valued  $F_{T}^{D(4)}(T4)$ in between, where
\be
\beta_{\pm}={1\over 2}\left(1\pm{1\over \sqrt{1+2\epsilon_L}}\right)
\label{eq:Beta}
\ee
and $\beta_{+}=0.79$ for $\epsilon_{L}=1$. Because the subleading
contributions $F_{L}^{D(3)}$ become larger as
$\beta$ decreases, cf. Eqs. (\ref{eq:phi2}) and (\ref{eq:phi1}), the
crossover is shifted to $\beta_{+} \sim 0.65$.

In Fig.~2 we show our results for $F_{2}^{D(3)}$ and its decomposition
into the transverse and longitudinal (assuming $\epsilon_{L}=1$),
and the twist-2 and twist-4 components
for $x_{\Pom}=1.33\ 10^{-3}$. Because of the inequality for the
excited mass, $M^{2}
\geq 4m_{c}^{2}$, open charm excitation contributes only
at $Q^{2} \geq 4m_{c}^{2}{\beta \over 1-\beta}$ and the charm threshold
effects are clearly visible in Fig.~2. As the suppression
$\propto {1\over m_{c}^{2}}$ of the open charm contribution
to transverse twist-2 SF is to a large extent compensated by the
large pQCD scale $\bar{Q}^{2}$, the charm abundance is substantial
even in $F_{T}^{D(3)}$. It is still larger in the longitudinal SF, in
which the flavour dependence only enters via the lower integration limit
in Eq. (\ref{eq:Gsqr}). The fractional contribution of open charm
to $F_{T}^{D(3)}(T4)$ is similar to that in $F_{L}^{D(3)}$, but
the charm threshold effects are smoother because $\Phi_{1}^{2} \propto
k^{2}$ which suppresses the near-threshold cross section. Apart
from the charm threshold effects, the twist-2
transverse SF is practically flat vs. $Q^{2}$.
Whereas for $\beta=0.65$ the longitudinal and transverse
twist-4 contributions cancel each
other, at $\beta=0.9$ the twist-4 contribution is very large.
Notice how scaling violations slow down the rate of decrease
of twist-4 structure functions at large $Q^{2}$. We find
good agreement between our results and the recent experimental data
on $F_{2}^{D(3)}$ at $x_{\Pom}=1.33\ 10^{-3}$
from the H1 collaboration \cite{H1F2D3}.
In Fig.~3 we show our results for the $x_{\Pom}$ dependence of
$F_{2}^{D}(x_{\Pom},\beta,Q^{2})$ in comparison with the H1 experimental
data. Our formalism describes excitation of the continuum
states, and the real comparison between
theory and experiment is justified only for masses above the $1S$ and
$2S$ vector meson excitation, $M^{2} \gsim 3$\,GeV$^{2}$, which
for $\beta =0.9$ implies $Q^{2}=M^{2}{\beta \over 1-\beta}
\gsim 20$-30 GeV$^{2}$. In the spirit of the exclusive/inclusive
duality \cite{GNZlong},
our results for the continuum in the vector meson excitation
region must be applicable to the experimental data in which the vector
meson peaks have been smeared out. With these reservations, our
calculations reproduce rather well the experimentally observed
$Q^{2}$ and $x_{\Pom}$ dependence of $F_{2}^{D(3)}(x_{\Pom},\beta,Q^{2})$
over the whole range of $Q^{2}$ and $x_{\Pom}$. Notice the steeper
$x_{\Pom}$ dependence at smaller $x_{\Pom}$, which is a characteristic
prediction of pQCD models of diffractive DIS \cite{GNZ95,Bartels,Levin}.
The accuracy of the experimental data on diffractive DIS is not yet
sufficient to test different models for gluon density in the proton.
%marco some comments more
Theoretical evaluations of the short-distance dominated $F_{L}^{D(3)}$
and $F_{T}^{D(3)}(T4)$ must be good to $\sim$ 10-20\% accuracy (due to
uncertainties in $G(x,k^2)$ in the
perturbative region),
whereas for the leading twist $F_{T}^{D(3)}$ (where one has a large
penetration into the low-$k^2$ region) a conservative estimate
for the accuracy is $\sim$ 20-30\% \cite{GNZ95}.

It is interesting to comment on the possible impact of diffractive
higher twist on the proton SF at small $x$. The
experimentally observed small-$x$ growth of the proton structure
function can not go indefinitely without running into conflict with
the unitarity. In general, one can decompose the experimentally measured
proton SF into the Born, alias the DGLAP, component and
the unitarity correction \cite{BaroneUnit,NZ94}:
\be
F_{2}(x,Q^{2})=F_{2}(DGLAP;x,Q^{2})+\Delta_{U}F_{2}(x,Q^{2})
\label{eq:Unit}
\ee
The standard DGLAP evolution, which is linear in parton densities,
is strictly applicable only to the non-unitarized SF
$F_{2}(DGLAP;x,Q^{2})$. The unitarization of the proton SF
is one of the open non-perturbative problems. The AGK rules suggest
(i=T,L,2)
\cite{AGK,BaroneUnit}
\be
\Delta_{U}F_{i}(x,Q^{2})= -\int_{{x\over x_{max}}}^{1}
 {d\beta \over \beta}
F_{i}^{D(3)}(x_{\Pom},\beta,Q^{2})\, ,
\label{eq:Fshad}
\ee
where $x_{max}=$0.05-0.1 is the threshold for diffractive DIS.
We are interested in the twist-4 component of  the unitarity correction
coming from the diffractive twist-4 component of $q\bar{q}$ excitation.
Because the integral (\ref{eq:Fshad}) is dominated by the
contribution from $\beta \sim 0.5$, one can put $\beta=0.5$ in the scaling
violations factor (\ref{eq:Gsqr}) and take it out of the integral
(\ref{eq:Fshad}). Then, after little algebra, one finds
\bea
 \Delta_{U}F_{i}(T4;x,Q^{2})={ e_{f}^{2} \over 90 Q^{2} B_{3\Pom}}
{\cal F_{G}}(x,0.5,Q^{2})
\left\{\begin{array}{ll}
~~3,~~& {\rm for}~~ $i=T$\\
-4,~~& {\rm for}~~ $i=L$\\
-1,~~& {\rm for}~~ $i=2$\\
\end{array}
\right.
\label{eq:ShadT}
\eea
which does not depend on $x_{max}$ as soon as $x \ll x_{max}$. It is
convenient to present the diffractive twist-4 effects as the correction
factor $(1-{Q_{4,i}^{2}\over Q^{2}})$ to the leading-twist structure
function. Our results for the so defined  $Q_{4,2}^{2}$
are shown in Fig.~4. We find that $Q_{4,2}^{2}$
rises with decreasing $x$ and increasing $Q^{2}$, but it is relatively
small because of the strong cancelation of the longitudinal and
transverse twist-4 contributions. It is interesting to notice that
diffractive twist-4 correction to the longitudinal structure function
of the proton is much larger,
\beq
Q_{4,L}^{2} = {4 \over R}Q_{4,2}^{2} \sim 20 Q_{4,2}^{2},
\label{eq:Q2twistL}
\eeq
where we use $R= \sigma_{L}/\sigma_{T} \approx 0.2$ \cite{NZHERA}
(for an indirect experimental estimates see \cite{FLHERA}).
\medskip\\
Of course, the above estimated diffractive higher twist from diffractive
DIS at large-$\beta$ is only one of the many possible sources of twist-4
corrections to the proton structure function at small $x$. We focused
on it for the reasons that it is parameter-free calculable in pQCD,
is enhanced by the gluon structure function of the proton squared
and receives a substantial contribution from the heavy flavour excitation.
At the moment, one can not exclude that a comparable contribution to
$Q_{4,i}^{2}$ will come from the small-$\beta$ diffraction.

\vspace{1.5cm}

\noindent{\large \bf  Summary and Conclusions.}
\bigskip

We have presented the calculation of the Born approximation for the higher
twist-4 corrections to the transverse diffractive structure function
of the proton. It is similar to the longitudinal diffractive structure
function which to the Born approximation is entirely twist-4. Both
twist-4 diffractive structure functions are short-distance dominated
and are perturbatively calculable. Their contribution
to the diffractive structure function is large and  dominates
$F_{2}^{D(3)}$ at $\beta \gsim 0.9$. Evidently, this twist-4 component
of $F_{2}^{D(3)}$ must be removed before venturing the DGLAP evolution
analysis of the experimentally measured diffractive structure function.
Because the pQCD calculation of the twist-4 corrections to
$F_{2}^{D(3)}$ is parameter free, such a correction of the diffractive
structure function must not cause any ambiguities.

Taking for guidance the AGK rules, we also evaluated the impact of
the diffractive twist-4 to the twist-4 terms in the proton structure
function. The estimate
(\ref{eq:Q2twistL}) serves as an warning that diffractive twist-4 corrections
to small-$x$ longitudinal
structure functions of the proton can be surprisingly large.
\newpage

% ======================================================================

% ======================================================================

\noindent {\Large \bf Figure captions }
\begin{enumerate}

\item[Figure 1] :
One of the 16 Feynman diagrams for diffractive cross section.

\item[Figure 2] :
The $Q^{2}$ dependence of the diffractive structure function 
and its decomposition into the transverse (T) and longitudinal (L)
components and into the twist-4 (T4) and twist-2 (T2) components.
The open charm threshold in $F_{2}^{D(3)}(x_{\Pom},\beta,Q^{2})$
is seen from the comparison of
solid curves for excitation of 4 flavours ($u+d+s+c$) and dotted
curves for excitation of light flavours ($u+d+s$). The predictions
are compared with the H1 experimental data on
$F_{2}^{D(3)}(x_{\Pom},\beta,Q^{2})$ at $x_{\Pom}=1.33\ 10^{-3}$ \cite{H1F2D3} for
$\beta=0.65$ (box $a$) and $\beta=0.9$ (box $b$).

\item[Figure 3] :
Our predictions for the $x_{\Pom}$ dependence of diffractive structure
function for $\beta=0.65$ (the two top raws) and $\beta=0.9$ (the
two bottom raws) in comparison with the H1 experimental data
\cite{H1F2D3} for several values of $Q^{2}$.

\item[Figure 4] :
The scale $Q_{4,2}^{2}$ for the AGK evaluation of the diffractive
higher twist contribution to the proton structure function at
small $x$.

\end{enumerate}

% =================================================================
\newpage 

\newpage

\begin{figure}[http]
  \begin{center}
    \mbox{\epsfig{file=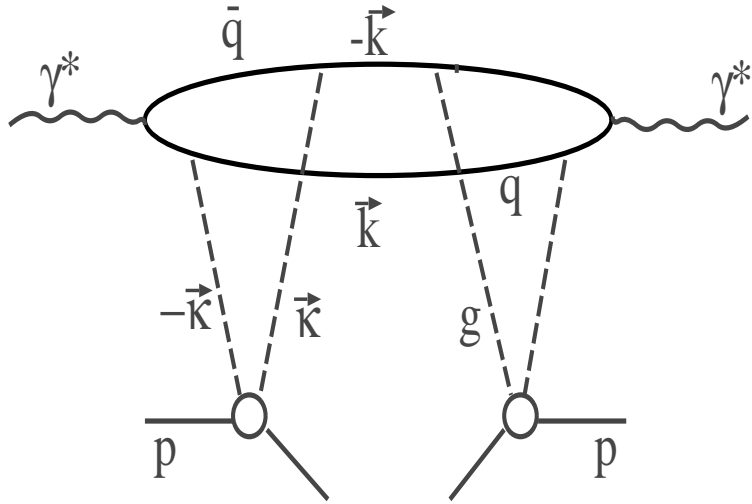,height=7cm,width=10cm,clip=}}
   \caption{One of the 16 Feynman diagrams for diffractive cross section.}
   \label{fig1}
\end{center}
\end{figure}

\begin{figure}[http]
  \begin{center}
    \mbox{\epsfig{file=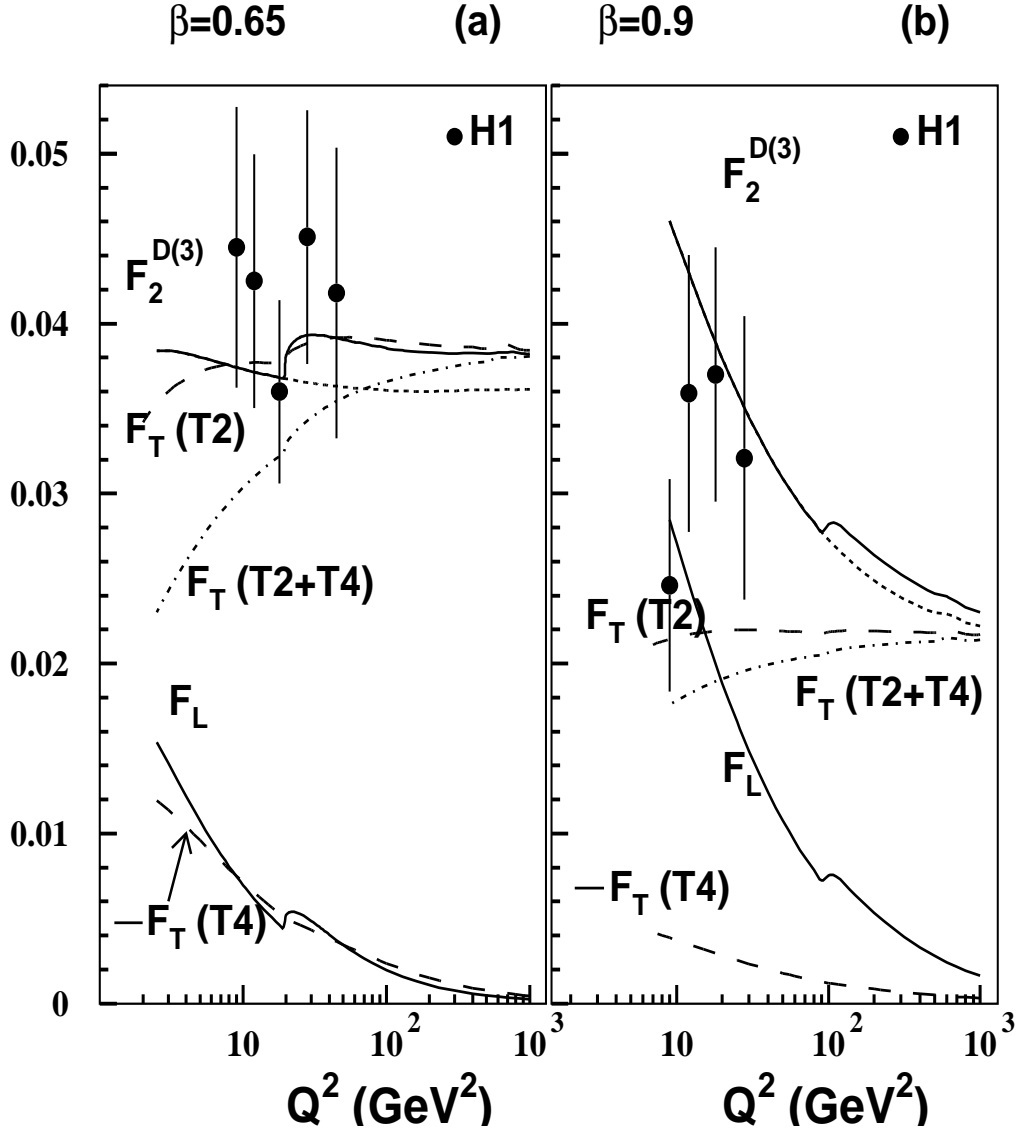,height=15cm,width=16cm,clip=}}
   \caption{The $Q^{2}$ dependence of the diffractive structure function
and its decomposition into the transverse (T) and longitudinal (T)
components and into the twist-4 (T4) and twist-2 (T2) components.
The open charm threshold in $F_{2}^{D(3)}(x_{\Pom},\beta,Q^{2})$
is seen from the comparison of
solid curves for excitation of 4 flavors ($u+d+s+c$) and dotted
curves for excitation of light flavors ($u+d+s$). The predictions
are compared with the H1 experimental data on
$F_{2}^{D(3)}(x_{\Pom},\beta,Q^{2})$ at $x_{\Pom}=1.33\ 10^{-3}$ \cite{H1F2D3} for
$\beta=0.65$ (box $a$) and $\beta=0.9$ (box $b$).}
   \label{fig:2ab}
\end{center}
\end{figure}

\newpage

\begin{figure}[http]
  \begin{center}
    \mbox{\epsfig{file=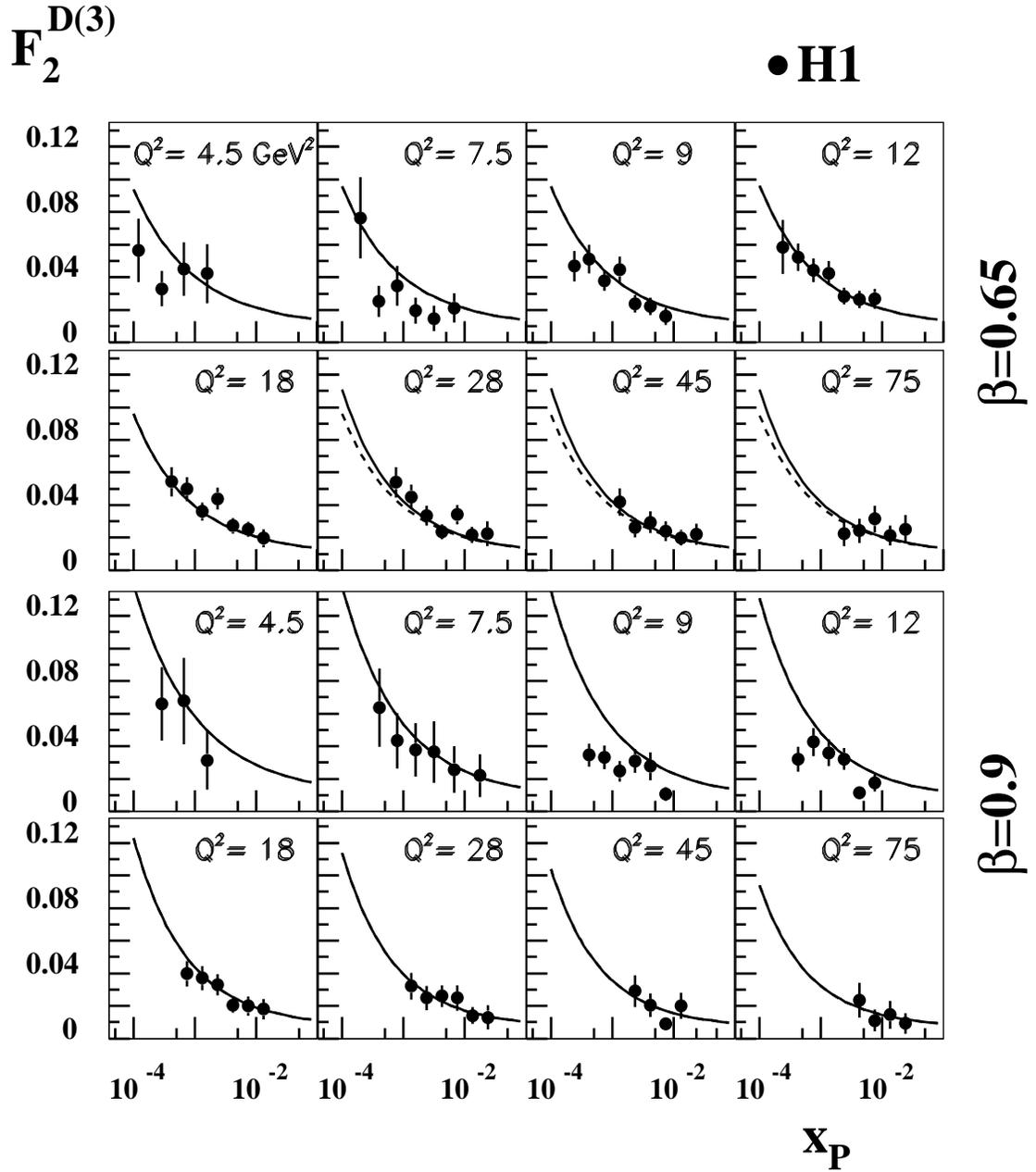,height=15cm,width=15cm,clip=}}
   \caption{Our predictions for the $x_{\Pom}$ dependence of diffractive structure
function for $\beta=0.65$ (the two top raws) and $\beta=0.9$ (the
two bottom raws) in comparison with the H1 experimental data
\cite{H1F2D3} for several values of $Q^{2}$.}
   \label{fig:3}
\end{center}
\end{figure}

\newpage

\begin{figure}[http]
  \begin{center}
    \mbox{\epsfig{file=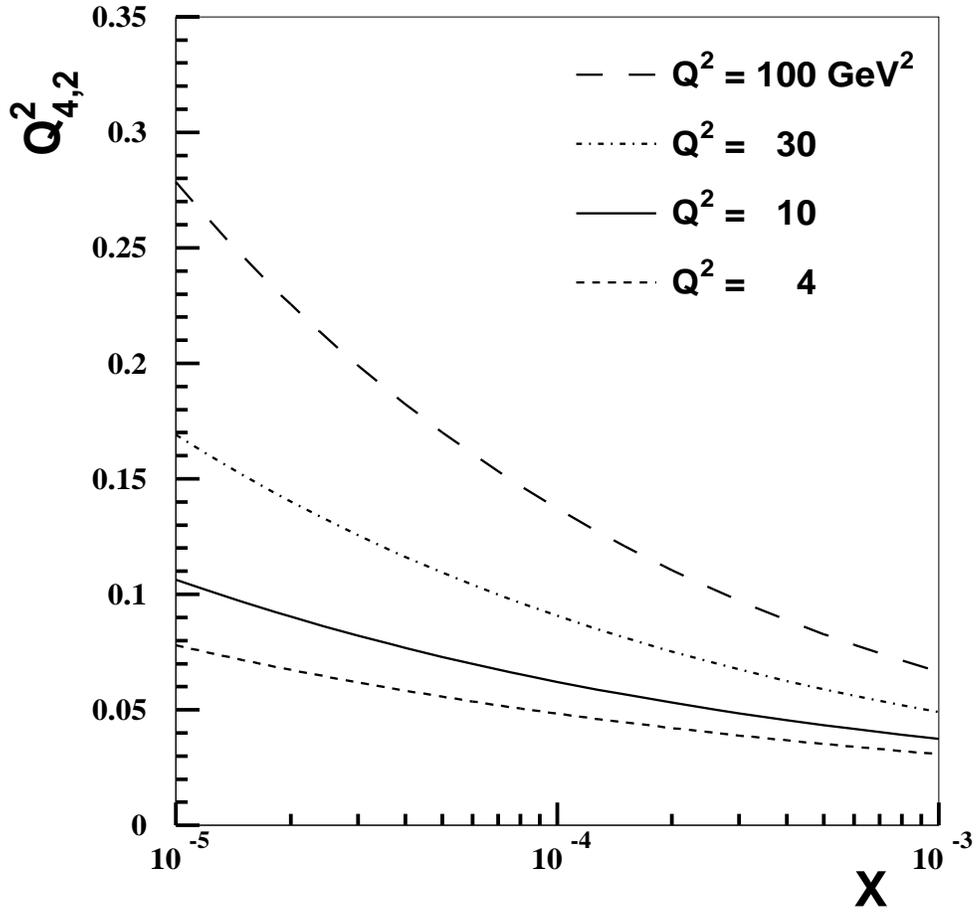,height=15cm,width=15cm,clip=}}
   \caption{The scale $Q_{4,2}^{2}$ for the AGK evaluation of the diffractive
higher twist contribution to the proton structure function at
small $x$.}
   \label{fig:4}
\end{center}
\end{figure}

\end{document}